%% file: Half-integer-arxiv.tex
\documentclass[12pt]{article}                  
\usepackage{jeosman}
\usepackage{amsmath}
\usepackage[subnum]{cases}
\usepackage{graphicx} 
\usepackage{subfigure} 
\usepackage{cite}
\usepackage{geometry}
\usepackage{color}
\usepackage{titlesec}
\usepackage{fancyhdr}
\usepackage{url}
\usepackage{soul}
\usepackage[frame,a4,center]{crop}

\begin{document}

\title{Half-integer order Bessel beams}

\author {Adrian Carbajal-Dominguez$^*$\thanks {adrian.carbajal@ujat.mx} }
\author {Jorge Bernal$^1$} 
\author{Alberto Mart\'in Ruiz$^2$}
\author{Gabriel Mart\'inez-Niconoff$^3$ }
\author{Jos\'e Segovia$^1$}

\address{$^1$\textit{Universidad Ju\'arez Aut\'onoma de Tabasco,\\ Divisi\'on Acad\'emica de Ciencias B\'asicas, Cunduac\'an, Tabasco, C.P. 86690, M\'exico}}
\address{$^2$\textit{Universidad Nacional Aut\'onoma de M\'exico,\\ Instituto de Ciencias Nucleares, C.P. 04510 M\'exico, D.F.,
 M\'exico }}
\address{$^3$\textit{Intituto Nacional de Astrof\'isica, \'Optica y Electr\'onica,\\ Apdo. postal 51 y 216, C.P. 7200, Puebla, Pue. 
M\'exico}}
\maketitle


\abstractname\\{Optical beams are solutions to the paraxial wave equation (PWE). In this work we report a new, to our knowledge, optical beam. We solve the PWE by using the angular spectrum of plane waves theory in circular cylindrical coordinates. This lead us to solutions expressed  in integral form that can be evaluated in closed-form for half-integer order Bessel functions. We show how to implement numerical simulation and how these results are confirmed by experiment. We believe this approach facilitates the analysis and synthesis of generic optical beams.}

\keywords {Bessel beams, paraxial wave equation}

\section{Introduction}

\input{introdnew2}

\section{Theory}
\subsection{Angular spectrum representation for PWE}

To begin with, it is well known that solutions to  the Helmholtz equation $\nabla^2\phi + k^2 \phi = 0$\cite{goodman_introduction_2005} in the form 
\begin{equation}
\phi(x,y,z)=f(x,y,z)exp(ikz)
\end{equation}
with the condition
\begin{equation}
\frac{\partial^2 f}{\partial z^2} << 2k\frac{\partial f}{\partial z}
\end{equation}

lead to the paraxial wave equation\cite{siegman_lasers_1986}

\begin{equation} \label{paraxial}
\nabla_{\bot}^2 f+i2k\frac{\partial f}{\partial z}=0
\end{equation}

where $\nabla_{\bot}^ 2$ is the transversal Laplacian operator and $k=2\pi/\lambda$, $\lambda$ is the wavelength.\\ 
Now, we search for solutions to PWE in the ASPW representation \cite{mandel_optical_1995}. In this representation solutions to Helmholtz equation are constructed as a superposition of coherent plane waves whose amplitude is weighted by a function called Angular Spectrum Function (ASF) noted  by $A(u,v)$. In Cartesian coordinates it has the form 

\begin{equation}\label{integral_form}
f(x,y,z)=\int_{-\infty}^{\infty}\!\int_{-\infty}^{\infty}\!A(u,v)\exp[i2\pi(xu+yv+zp)]dudv .
\end{equation}

The $A(u,v)$ function is related to the amplitude transmittance $t(x,y)$ located at $z=0$ via its Fourier transform.
\begin{equation}
A(u,v)=\int_{-\infty}^{\infty}\!\int_{-\infty}^{\infty}\!t(x,y)\exp[-i2\pi(xu+yv)]dxdy .
\end{equation}
Now, we apply the same ideas to PWE. By substituting Eq.(\ref{integral_form}) into Eq.(\ref{paraxial}) one gets a solution of the form
\begin{equation}\label{ASPW-paraxial}
f(x,y,z)=\int_{-\infty}^{\infty}\!\int_{-\infty}^{\infty}\!A(u,v)\exp[-i \pi \lambda z (u^2+v^2)] \exp[i2\pi(xu+yv)]dudv. 
\end{equation}

Eq.(\ref{ASPW-paraxial}) can be transformed to circular cylindrical coordinates by means of suitable variable change. Invoking Jacobi-Anger identity \cite{abramowitz_handbook_1972} one obtains
\begin{equation}\label{aspw-cylindrical}
f(r,\theta,z)=\displaystyle\sum_{n=-\infty}^\infty i^n \exp(in\theta) \int_{0}^{2\pi}\!\int_{0}^{\infty} A(\rho, \psi)J_n(2\pi r \rho)\exp(-in\psi)exp(-i \pi \lambda z \rho^2)\rho d \rho d\psi
\end{equation}

The resulting ASF $A(\rho, \psi)$ depends on radial and angular coordinates in the integral.  
In order to facilitate the integration on the angular coordinate, we consider a separable angular spectrum with a simple angular dependence as in the form
\begin{equation}
A(\rho,\psi)=B(\rho)\exp(im\psi)\label{AS-paraxial},
\end{equation}
with $m \neq 0$. Performing the integration on the angular coordinate leads to the expression
\begin{equation}\label{integral-solution}
f(r,\theta,z)= 2\pi i^m \exp(im \theta) \int_{0}^{\infty}\! B(\rho)J_m(2\pi r \rho)exp(-i \pi \lambda z \rho^2)\rho d \rho.
\end{equation}
The angular spectrum in the form given by Eq.(\ref{AS-paraxial}) permits to obtain solutions to the PWE by finding $B(\rho)$ functions that integrate Eq.(\ref{integral-solution}) in closed-form. In fact, there are two integrals to be made. One for the plane boundary condition when $z=0$ and the other for any observation distance with $z > 0$. 

\subsection{Half-integer order Bessel functions solutions}
In order to obtain the corresponding angular spectrum function $B(\rho)$, one has to evaluate Eq.(\ref{integral-solution}) at $z=0$. This integral is evaluated in closed-form when \cite{gradshteyn_table_1994} 
\begin{equation} \label{half-order-aspw}
B(\rho)=J_{n/2}(a^2 \rho^2),
\end{equation}
where $a$ is a parameter; $n$ an integer; $J_{n/2}$ are half-integer order Bessel functions. Replacing Eq.(\ref{half-order-aspw}) in Eq.(\ref{integral-solution}) and integrating for $z=0$  leads to the result 

\begin{equation}\label{half-order-t}
t(r,\theta)=f(r,\theta,z=0)=\frac{\pi}{a^2}i^m J_{m/2}\left(\frac{\pi^2 r^2}{a^2}\right)\exp(im\theta).
\end{equation}
In Eq.(\ref{half-order-t}) $t(r,\theta)$ is the amplitude transmittance associated to the angular spectrum function considered in Eq.(\ref{half-order-aspw}). This will be a very important result for the generation of these beams by diffraction. 


Continuing with Eq. (\ref{integral-solution}) for $z > 0$, we can say it describes the propagation of the optical field in the $z$ coordinate. In order to perform this integral, we consider the results \cite{gradshteyn_table_1994}

\begin{equation}\label{identy-1}
\int_{0}^{\infty} x \cos(\alpha x^ 2)J_{n/2}(\beta x^ 2)J_n(2\gamma x)dx= \frac{1}{2\sqrt{\beta^2-\alpha^2}}\cos\left(\frac{\alpha \gamma^2}{\beta^2-\alpha^2}\right) J_{n/2}\left(\frac{\beta \gamma^2}{\beta^2-\alpha^2}\right)
\end{equation}
and
\begin{equation}\label{identy-2}
\int_{0}^{\infty} x \sin(\alpha x^ 2)J_{n/2}(\beta x^ 2)J_n(2\gamma x)dx= \frac{1}{2\sqrt{\beta^2-\alpha^2}}\sin\left(\frac{\alpha \gamma^2}{\beta^2-\alpha^2}\right) J_{n/2}\left(\frac{\beta \gamma^2}{\beta^2-\alpha^2}\right)
\end{equation}
for $0<\alpha<\beta$.
\\
By writing Eq.(\ref{identy-1}) and (\ref{identy-2}) in complex form and combining them with the integral (\ref{integral-solution}) one obtains   
\begin{equation}
f(r,\theta ,z)=\pi i^m \exp(im\theta)\frac{1}{Q(z)}\exp\left(-i\frac{\pi^3 z\lambda r^2}{a^4-\pi^2 r^2}\right)J_{m/2}\left(\frac{a^2 \pi^2 r^2}{Q(z)^2}\right)
\end{equation}
with $Q(z)=\sqrt{a^4-\pi^2\lambda^2z^2}$. 
This is an exact solution to PWE and it represents an optical beam whose irradiance is given by
\begin{equation}
I=\frac{\pi^2}{Q(z)^2}J_{m/2}^2\left(\frac{a^2\pi^2 r^2}{Q(z)^2}\right)
\end{equation}
This is, to our knowledge, a new kind of optical beam and we propose to call it \textit{Half-order Bessel beam}. It is also shown that such an optical beam is simply generated by the diffraction produced by the amplitude transmittance given in Eq.(\ref{half-order-t}) when illuminated with a plane wave. 
Fractional-order Bessel functions are known to be solutions to inhomogeneous Helmholtz equation and they are useful to approximate anlytical functions \cite{jung_approximation_2012}. Despite other authors \cite{Bandres2008b} have reported general solutions for paraxial beams in circular coordinates, they do not include the family of solutions  we are reporting here.

\section{Numerical simulation and experimental synthesis}
One interesting feature of the present approach is that the amplitude transmittance is already calculated. Hence, it is straightforward to implement a numerical simulation for the diffracted field. For simplicity, we only consider the transmittance real part.

\begin{figure}[htbp]
\centering
\subfigure[][]{\includegraphics[width=30mm]{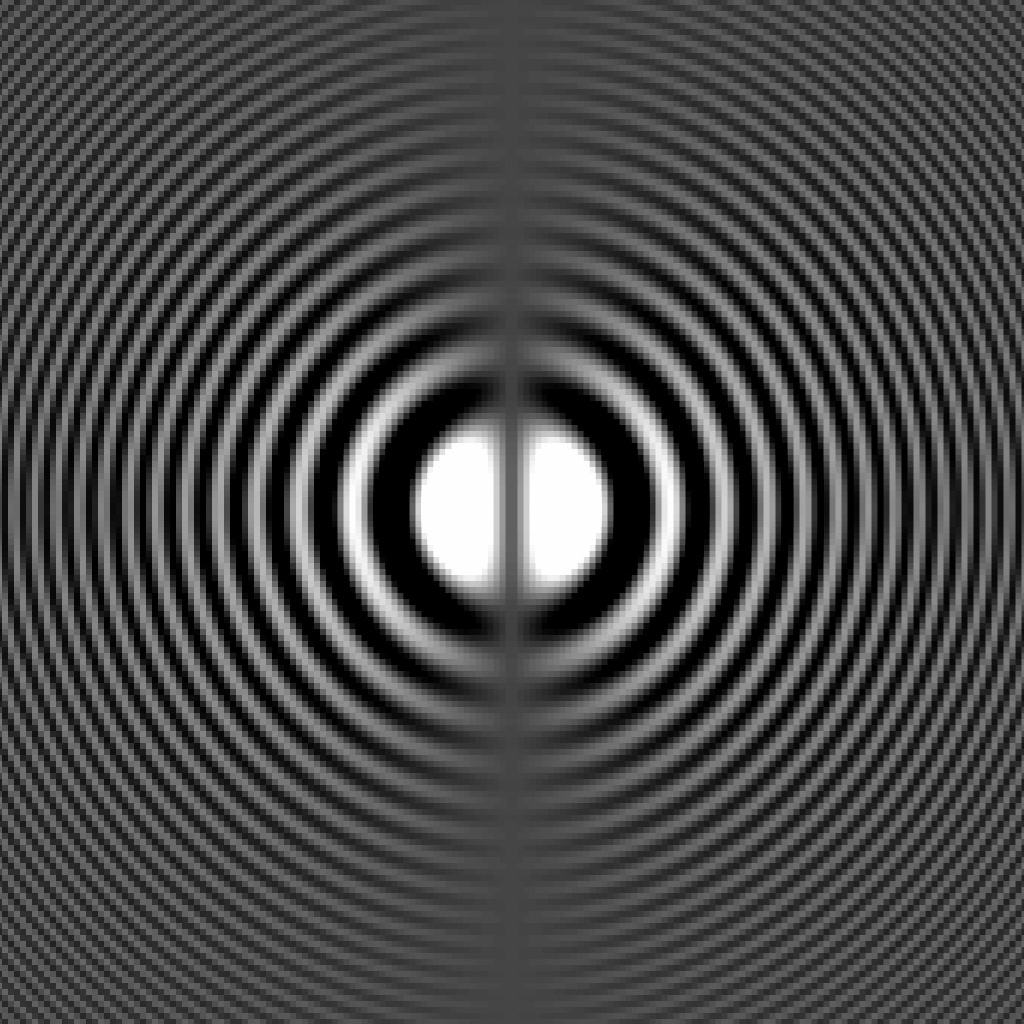}}\hspace{8pt}%
\subfigure[][]{\includegraphics[width=30mm]{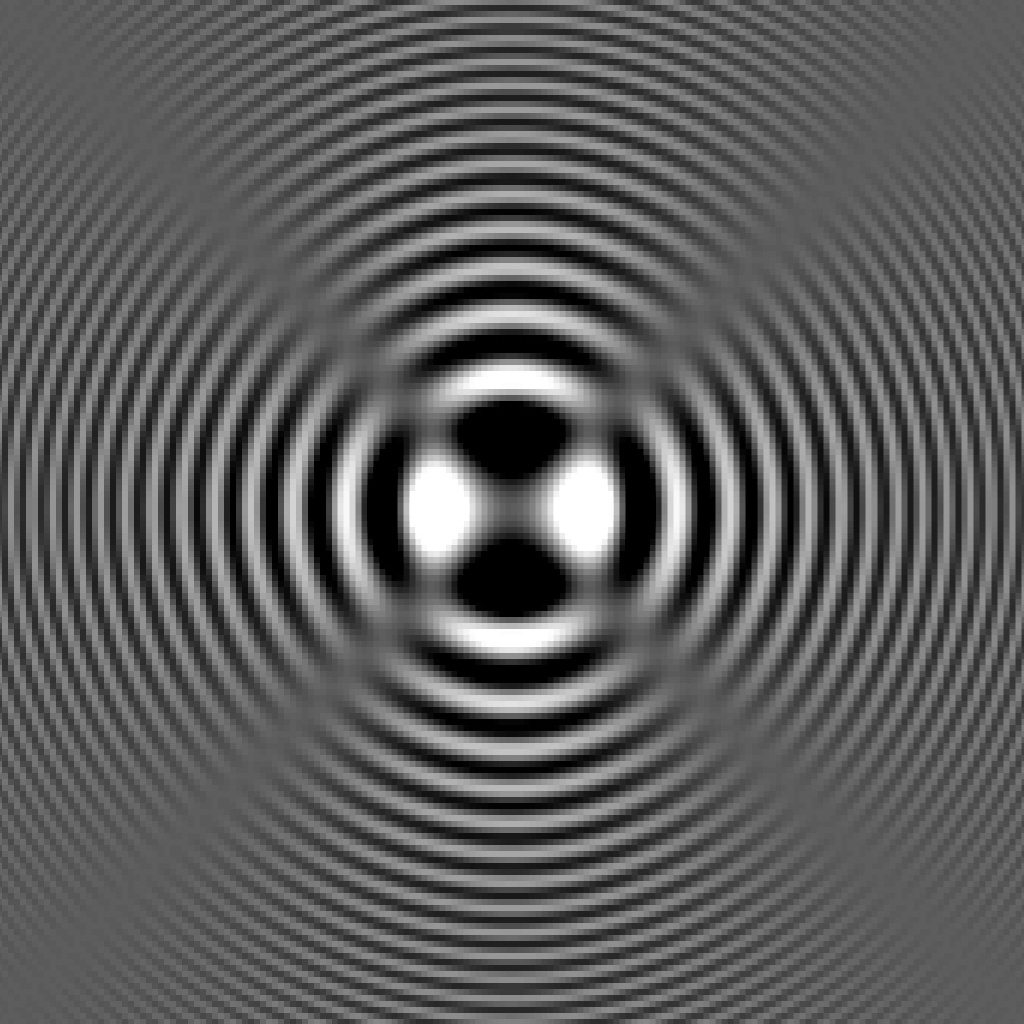}}\hspace{8pt}%
\subfigure[][]{\includegraphics[width=30mm]{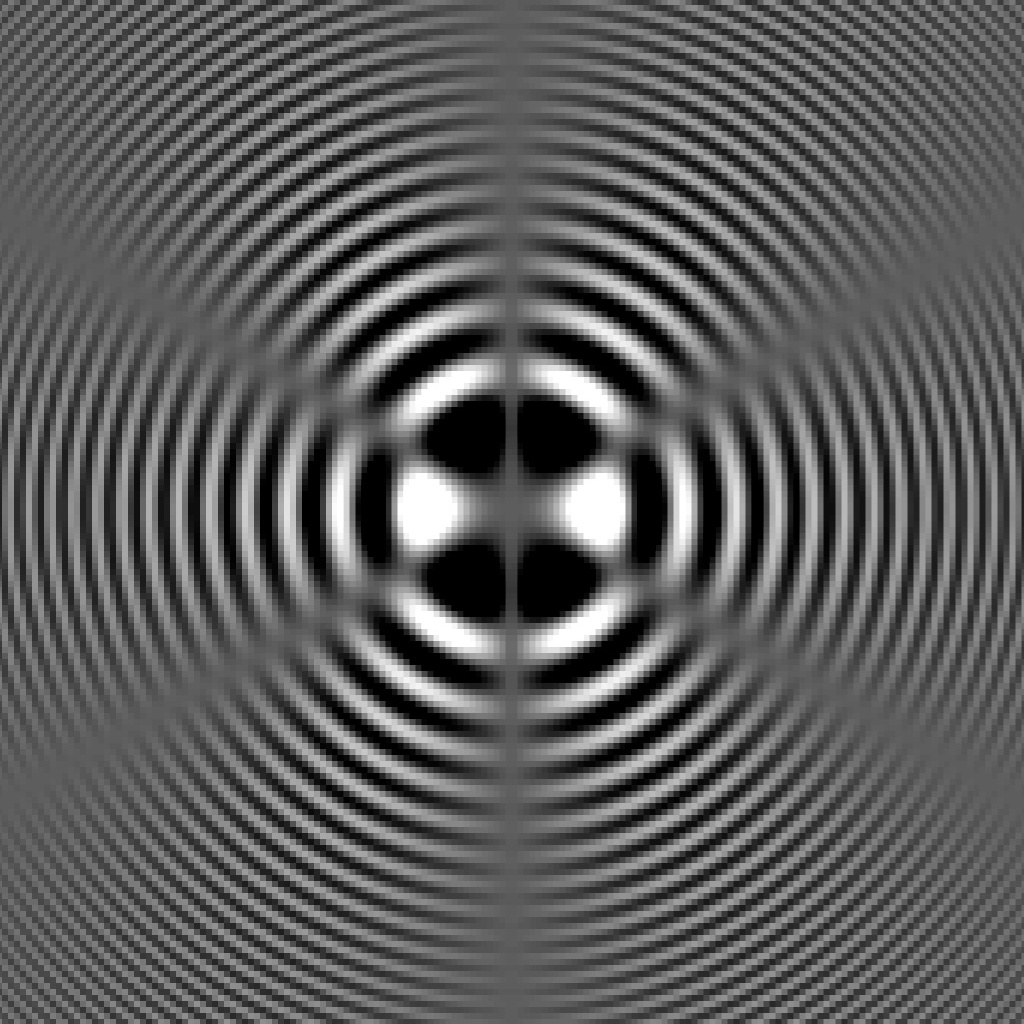}}


\subfigure[][]{\includegraphics[width=30mm]{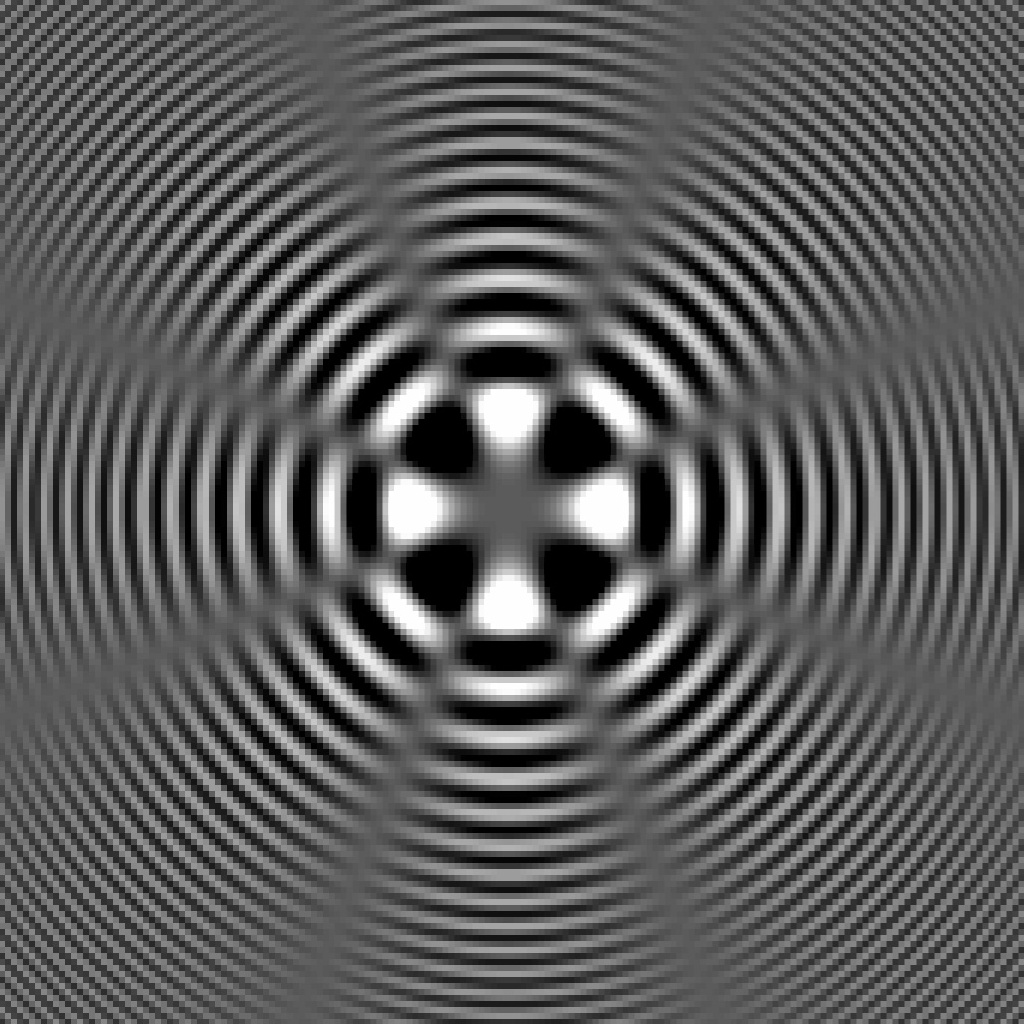}}\hspace{8pt}%
\subfigure[][]{\includegraphics[width=30mm]{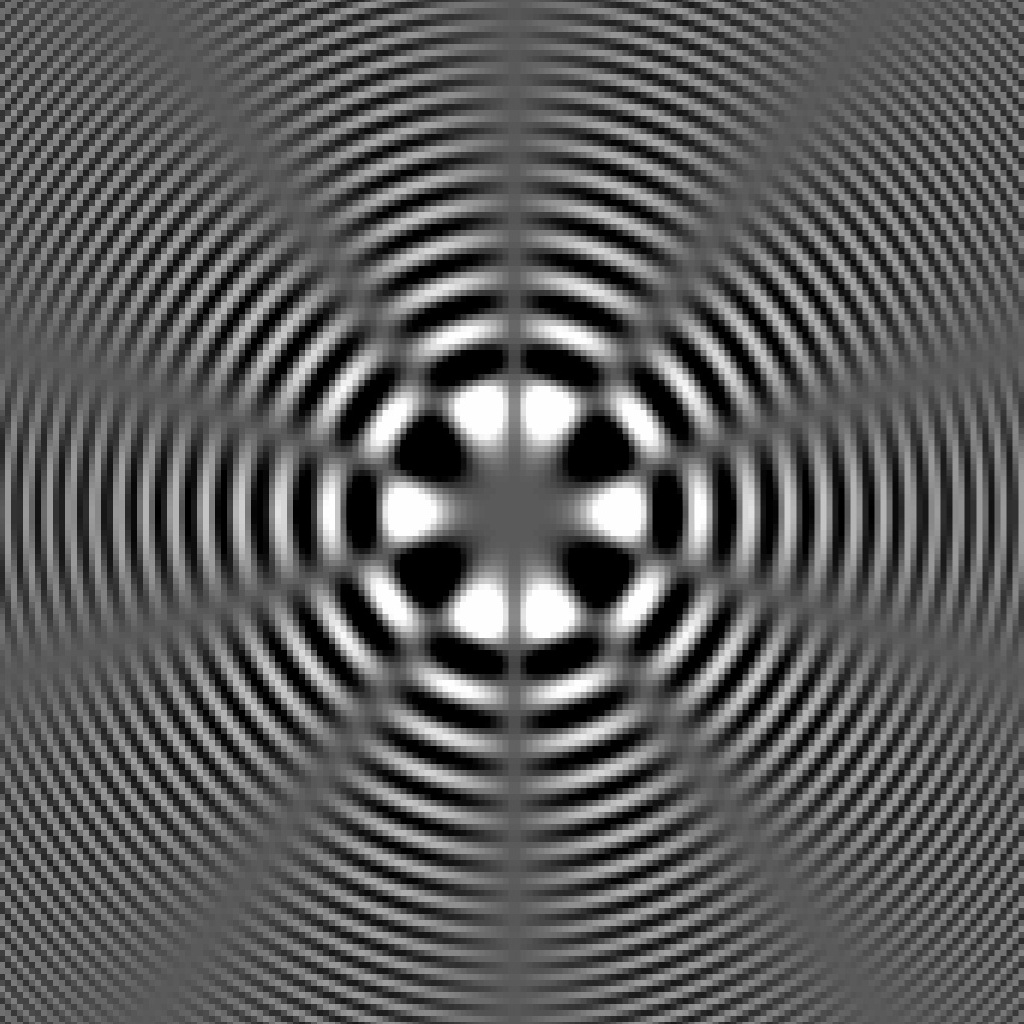}}\hspace{8pt}%
\subfigure[][]{\includegraphics[width=30mm]{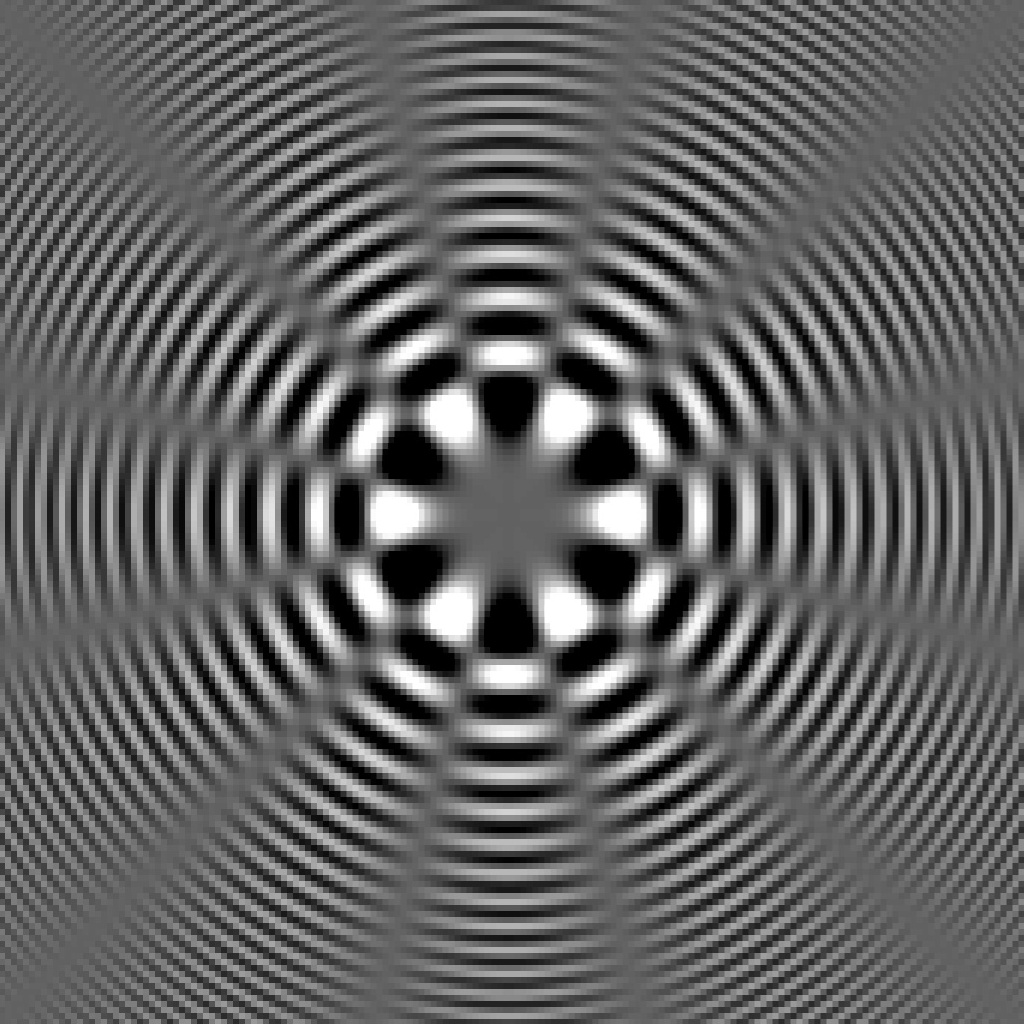}}
\caption{Transmittances functions for the six first solutions. (a) (c) $m$ = 1 to 3. (d) (f) $m$ = 4 to 6.}%
\label{fig:transmittances}
\end{figure}

The Eq.(\ref{half-order-t}) real part can be represented as squared matrix denoted by $T$ and $\xi,\zeta$ matrix indices. Then, matrix elements are defined as
 \begin{equation}\label{transmitacia-matrix}
 \textbf{T}_m^{\xi,\zeta}=J_{m/2}(\sigma \sqrt{\xi^2+\zeta^2})\cos[m\epsilon \arctan(\zeta / \xi)]
 \end{equation}
 with $\sigma$ and $\epsilon$ parameters. This matrix is scaled to 256 gray levels and it is saved as a bitmap image. Examples of the resulting images form $m = 1$ to 6 are shown in Fig. (\ref{fig:transmittances}).   
 
For the numerical simulation for the diffracted field we consider Eq.(\ref{ASPW-paraxial}) in matrix form. We define the matrix 
\begin{equation}
\textbf{A}=FFT\left\lbrace \textbf{T}_m \right\rbrace
\end{equation} 
where FFT is the Fast Fourier transform. We also define the matrix associated to the optical field evolution, whose elements are given in the form
\begin{equation}
\textbf{C}_{\xi,\zeta}(z)=\exp[-i \pi \lambda z(\xi^2+\zeta^2)].
\end{equation}

Hence, the resulting diffracted field for a given $z$ is 
\begin{equation}\label{field-num}
\textbf{F}_{(z)}=FFT^{-1}\lbrace \textbf{A}\cdot \textbf{C}(z)\rbrace
\end{equation}
where $FFT^{-1}$ is the inverse fast Fourier transform. Eq.(\ref{field-num}) allows to calculate the diffracted optical field by means of FFT which is available in most mathematical software. Care must be taken of properly accommodate the high and low FFT frequencies. Most programs include centring functions related to FFT functions.\\
  
\begin{figure}[htbp]
\centering
{\includegraphics[height=70mm]{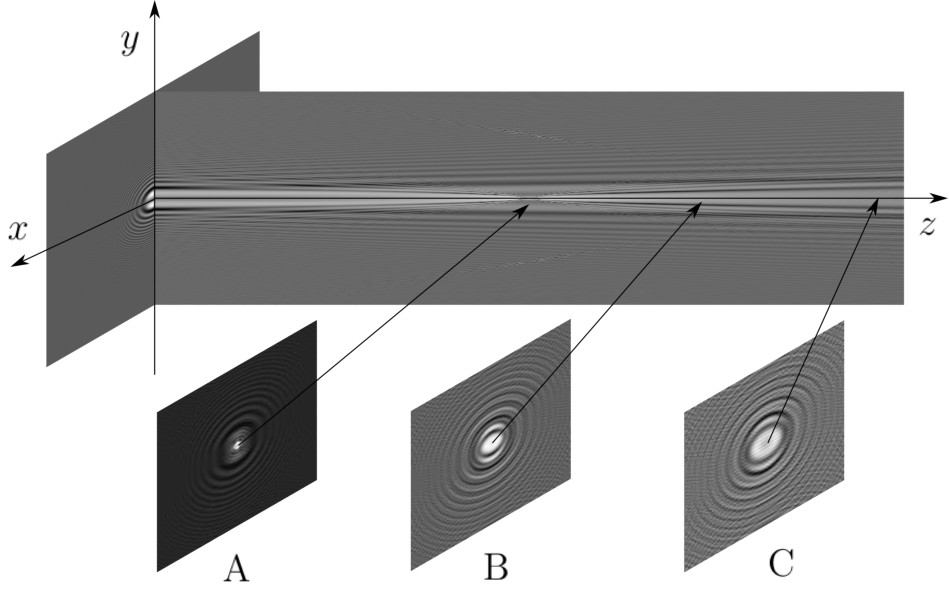}}
\vspace{22pt}
\caption{Numerical results example for $m =$1. Propagation at different $z$ distances are computed. A waist is observed in A, while B-C show the transversal optical field for this case. }
\label{fig:numerical-results}
\end{figure}

The results of numerical calculations of the diffracted field for several $z$ values of  are shown in Fig. \ref{fig:numerical-results}. Light is propagated in the positive z direction, from left to right. One can see that a focusing region appears. Most of the energy is concentrated around the optical axis, as expected for any solution to PWE. Several solutions were studied, and all propagate in the same manner.\\ 

\begin{figure}[htbp]
\centering
\subfigure[][]{\includegraphics[width=30mm]{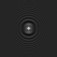}}\hspace{8pt}%
\subfigure[][]{\includegraphics[width=30mm]{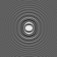}}\hspace{8pt}%
\subfigure[][]{\includegraphics[width=30mm]{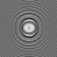}}

\subfigure[][]{\includegraphics[width=30mm]{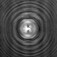}}\hspace{8pt}%
\subfigure[][]{\includegraphics[width=30mm]{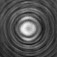}}\hspace{8pt}%
\subfigure[][]{\includegraphics[width=30mm]{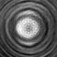}}

\subfigure[][]{\includegraphics[width=30mm]{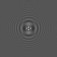}}\hspace{8pt}%
\subfigure[][]{\includegraphics[width=30mm]{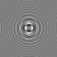}}\hspace{8pt}%
\subfigure[][]{\includegraphics[width=30mm]{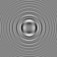}}

\subfigure[][]{\includegraphics[width=30mm]{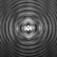}}\hspace{8pt}%
\subfigure[][]{\includegraphics[width=30mm]{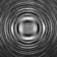}}\hspace{8pt}%
\subfigure[][]{\includegraphics[width=30mm]{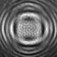}}

\subfigure[][]{\includegraphics[width=30mm]{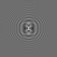}}\hspace{8pt}%
\subfigure[][]{\includegraphics[width=30mm]{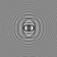}}\hspace{8pt}%
\subfigure[][]{\includegraphics[width=30mm]{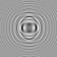}}

\subfigure[][]{\includegraphics[width=30mm]{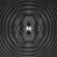}}\hspace{8pt}%
\subfigure[][]{\includegraphics[width=30mm]{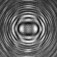}}\hspace{8pt}%
\subfigure[][]{\includegraphics[width=30mm]{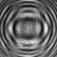}}
\caption{Numerical and Experimental results for $m=1, 2, 3$. a)- c) Numerical results, d)- f) experimental results for $m = 1$. g)- i) Numerical results, j)- l) experimental results for $m = 2$. m)- o) Numerical results, p)- r) experimental results for $m = 3$.}%
\label{fig:results-1-3}
\end{figure}

\begin{figure}[htbp]
\centering
\subfigure[][]{\includegraphics[width=30mm]{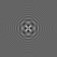}}\hspace{8pt}%
\subfigure[][]{\includegraphics[width=30mm]{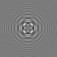}}\hspace{8pt}%
\subfigure[][]{\includegraphics[width=30mm]{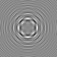}}

\subfigure[][]{\includegraphics[width=30mm]{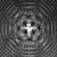}}\hspace{8pt}%
\subfigure[][]{\includegraphics[width=30mm]{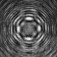}}\hspace{8pt}%
\subfigure[][]{\includegraphics[width=30mm]{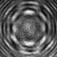}}

\subfigure[][]{\includegraphics[width=30mm]{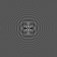}}\hspace{8pt}%
\subfigure[][]{\includegraphics[width=30mm]{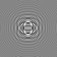}}\hspace{8pt}%
\subfigure[][]{\includegraphics[width=30mm]{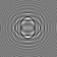}}

\subfigure[][]{\includegraphics[width=30mm]{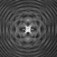}}\hspace{8pt}%
\subfigure[][]{\includegraphics[width=30mm]{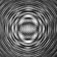}}\hspace{8pt}%
\subfigure[][]{\includegraphics[width=30mm]{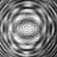}}

\subfigure[][]{\includegraphics[width=30mm]{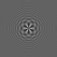}}\hspace{8pt}%
\subfigure[][]{\includegraphics[width=30mm]{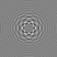}}\hspace{8pt}%
\subfigure[][]{\includegraphics[width=30mm]{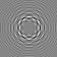}}

\subfigure[][]{\includegraphics[width=30mm]{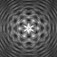}}\hspace{8pt}%
\subfigure[][]{\includegraphics[width=30mm]{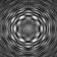}}\hspace{8pt}%
\subfigure[][]{\includegraphics[width=30mm]{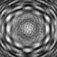}}
\caption{Numerical and Experimental results for $m=4, 5, 6$. a)- c) Numerical results, d)- f) experimental results for $m = 4$. g)- i) Numerical results, j)- l) experimental results for $m = 5$. m)- o) Numerical results, p)- r) experimental results for $m = 6$.}%
\label{fig:results-4-6}
\end{figure}

For the experimental synthesis we construct the transmittance in the form of gray level image printed on photographic film. Diffraction from film slides were studied under coherent plane wave illumination. The diffracted optical field was then recorded at different observation distances using a CCD camera without a lens. Results are shown for $m = 1, 6$ in figures \ref{fig:results-1-3} and \ref{fig:results-4-6} for propagation distances schematically depicted in Fig. \ref{fig:numerical-results} for A, B and C characteristic propagation distances.

For each case, one can see there is a good agreement between the proposed numerical method and the experimental observation, as shown in Fig. \ref{fig:results-1-3} and Fig. \ref{fig:results-4-6}. Experimental results show that these solutions effectively maintain generic characteristics
of optical beams. For instance, they are able to keep most of their energy around optical axis. In the near-field region where $z \sim 0$, the optical field appears to be almost invariant. Then it evolves towards a focalization region. Afterwards, the emerging optical field propagates with an almost constant slightly divergent transverse profile.\\
Nevertheless, there differences among our experimental  with numerical results. For instance, for cases when $m = 4$ and $m = 6$ in Fig. \ref{fig:results-4-6} a)-d) and m)-p). We believe these differences are due to phase terms that are discarded in scalar theory and that correspond to vectorial  behaviour.\\

\section{Conclusions}
We introduced an ASPW representation for the PWE. Using it, one can obtain some solutions to the PWE by  exact evaluation of the diffraction integrals in the ASPW context. We have showed that a separable angular spectrum facilitates the integration for the radial and angular coordinates. The angular coordinate is introduced in a form that is easily evaluated, and the entire problem reduces to finding the integral in the radial coordinate. We showed that one family of solutions for the radial coordinate is the functions called half-integer order Bessel functions with a quadratic argument. The ASPW treatment for the PWE also proves to be very helpful in  numerical calculations of the diffracted field. Accurate results are obtained using FFT. The numerical and experimental results are in good agreement and show that these solutions preserve most of their energy around the optical axis, like any paraxial optical beam.  
A.C.-D, J.B. and J.S.  want to thank Universidad Ju\'arez Aut\'onoma de Tabasco for the support for this work.


\bibliographystyle{jeos}

\end{document}

%% file: introdnew2.tex
It is well known that optical beams are solutions to paraxial wave equation (PWE)\cite{siegman_lasers_1986}. They are able to maintain most of their energy homogeneously distributed very close to the propagation axis. Their particular features are very attractive to many branches of optical sciences and they have found an important number of applications such as information transmission in free space\cite{majumdar_free-space_2010}, optical trapping\cite{arlt_optical_2001}, nano-optics\cite{courvoisier_surface_2009}, microscopy \cite{Fahrbach2010}.\

In a classical work, Durnin \cite{durnin_exact_1987} obtained the $J_0$ Bessel beam by expressing a solution to Helmholtz equation in integral form which turned out to be evaluated in closed-form giving as result such optical beam.
 
Since then, several optical beams have been reported in literature. They usually are expressed in terms of special functions.  
It is worth mentioning the examples of Bessel\cite{durnin_exact_1987}, Gaussian\cite{siegman_lasers_1986-2}, Bessel-Gaussian\cite{gori_bessel-gauss_1987}, Laguerre-Gaussian, Hermite-Gaussian\cite{Zauderer1986}, Airy\cite{siviloglou_observation_2007,ellenbogen_nonlinear_2009},Mathieu\cite{Gutierrez-Vega2000}, Pearcey\cite{ring_auto-focusing_2012}, accelerated beams\cite{Bandres2009}, Helmholtz-Gauss \cite{gutierrez-vega_helmholtz-gauss_2005}, Mathieu \cite{gutierrez-vega_experimental_2001}, Gauss-Laguerre \cite{piestun_propagation-invariant_2000}, Hankel-Bessel \cite{kotlyar_hankel-bessel_2012}, Hermite and Hypergeometric \cite{kotlyar_hypergeometric_2007} optical beams.\\

Authors have explored different theoretical routes to obtain such optical beams, ranging from the direct solution of PWE, to ansatz constructed from properly tailored functions \cite{Bandres2008b}. As in Durnin case, solutions to wave equation have also turned out to be solutions to PWE. Others have been derived from diffraction theory. Rayleigh-Sommerfeld diffraction integral \cite{Torres-Vega1996}, angular spectrum representation \cite{Agrawal1979}, Huygens-Fresnel principle \cite{Williams1973}. It is worth mentioning accelerating beams whose rather particular behaviour was predicted by Berry and Balasz by analysing Schr\"oedinger equation \cite{Berry1979}.      
 
Every new paraxial solution has to be tested by numerical simulations which normally are more dependent on algorithms for special function calculation than they are on diffraction theory \cite{gutierrez-vega_helmholtz-gauss_2005_1}.    
Finally, optical beams need to be generated experimentally. Due to their intrinsic complexity, the preferred method of synthesis, in many cases, is holography \cite{Vasara1989,Rodrigo2011}. 

In this work we use a Durnin-like approach to solve the paraxial wave equation. We obtain a general solution by using the angular spectrum of wave  (ASPW) theory {\cite{goodman_introduction_2005, mandel_optical_1995}} in circular cylindrical coordinates and manage to perform the integral in closed-form by introducing a half-integer Bessel function with quadratic argument.
The numerical study of this solution is based on 2D fast Fourier transform (2D-FFT) inherited from the integral representation here proposed.   
We were able to produce this solutions experimentally simply by the diffraction of a prescribed function on a plane serving as a amplitude transmittance.
We believe our approach may be very useful to obtain  new solution to paraxial equation in closed-form or, for other more general cases, to numerically and experimentally produce new paraxial beams in simpler manner.